\documentclass[aps,twocolumn]{revtex4}
\usepackage[utf8]{inputenc}
 \usepackage{amsmath,amssymb}
\usepackage{color}
\begin{document}
\title{Corrections to Friedmann equations inspired by Kaniadakis entropy}
\author{Ahmad Sheykhi\footnote{asheykhi@shirazu.ac.ir}}
\address{Department of Physics, College of
Science, Shiraz University, Shiraz 71454, Iran\\
Biruni Observatory, College of Science, Shiraz University, Shiraz
71454, Iran}

 \begin{abstract}
Adopting the thermodynamics-gravity conjecture, and assuming the
entropy associated with the apparent horizon of the
Friedmann-Robertson-Walker (FRW) universe has the form of the
generalized Kaniadakis entropy, we extract the modified Friedmann
equations describing the evolution of the universe using the first
law of thermodynamics on the apparent horizon. We then investigate
the validity of the generalized second law of thermodynamics for
the universe enclosed by the apparent horizon.

\end{abstract}
 \maketitle

 \newpage
\section{Introduction\label{Intro}}
Inspired by the laws of black holes mechanics, the profound
connection between the laws of gravity and the first law of
thermodynamics was established \cite{Jac}. This connection is
usually called thermodynamics-gravity conjecture/correspondence
and is now generally accepted from theoretical point of view. The
deep connection between gravity and thermodynamics has been well
established at three levels. At the first level, it was argued
that the field equations of gravity can be written in the form of
the first law of thermodynamics on the horizon and vice versa (see
\cite{Jac,Pad1,Pad2,Pad3,CaiKim,Cai2,Shey1,Shey2,SheyCQ,SheyLog,SheyPL}
and references therein). At the second level, it was argued that
the field equations of gravity can be derived from statistical
mechanics. In this approach, by starting from the first
principles, namely the holographic principle and the equipartition
law of energy on the horizon degrees of freedom, one can obtain
the gravitational field equations \cite{Ver}. This approach is
known as entropic force scenario, which states that gravity is not
a fundamental force and can be emerged from the change in the
information of the system. This scenario has attracted a lot of
attentions (see e.g. \cite{Cai5,sheyECFE,ECEE,Visser,SRM}). At the
third level, it was argued that the spatial expansion of the
universe can be understood as the consequence of the emergence of
space. Thus, there is no pre-exist geometry or spacetime, and the
\textit{cosmic space emerges as the cosmic time progress}
\cite{PadEm}. Emergence scenario of gravity has been generalized
to Gauss-Bonnet, Lovelock and braneworld frameworks
\cite{CaiEm,Yang,Sheyem,Sheyem2,Sheyem3}.

It is important to note that, regardless the three approaches
mentioned above, in order to extract the gravitational field
equations from the thermodynamic arguments, the entropy expression
plays a crucial role. Any modification to the entropy expression
modifies the corresponding field equations of gravity
\cite{CaiLM,SheT1,Odin1,SheT2,Emm2,SheB1,SheB2,Odin2,Odin3,Odin4,Odin5}.

In this Letter we would like to study the effects of the
generalized Kaniadakis entropy on the cosmological field
equations. Modified Friedmann equations based on the generalized
Kaniadakis entropy was already explored in \cite{Lym}. Starting
from the relation $-dE=TdS $ on the apparent horizon of FRW
universe, the influence of the Kaniadakis entropy was explored
\cite{Lym}. Note that here $-dE$ is the energy flux crossing the
horizon within an infinitesimal period of time $dt$, while $T$ and
$S$ are, respectively, the temperature and the entropy associated
with the apparent horizon \cite{Lym}. It was shown that the
modified Friedmann equations contain new extra terms that
constitute an effective dark energy sector depending on the
Kaniadakis parameter $K$. It was also argued that the dark energy
equation of state parameter deviates from standard $\Lambda$CDM
cosmology at small redshifts, and remains in the phantom regime
during the history of the universe \cite{Lym}. Our work differs
from \cite{Lym} in that we modify the geometry part of the
cosmological field equations, and we assume the energy/matter
content of the universe is not affected by the generalized
Kaniadakis entropy. We believe this is more reasonable, since
entropy is a geometrical quantity and any modification to it
should change the geometry part of the field equations. In
addition, since our universe is expanding, thus we consider the
work term (due to the volume change) in the first law of
thermodynamics and write it as $dE=TdS+WdV$. Cosmological
implications of the modified Friedmann equations based on
generalized Kaniadakis entropy, and its influences on the early
baryogenesis and primordial Lithium abundance have been
investigated in \cite{Luci}. Other cosmological consequences of
the Kaniadakis entropy have been carried out in
\cite{Mor,Her,Dre,Kum1,Kum2}.

The structure of this Letter is as follows. In section \ref{Kan},
we review the origin of the Kaniadakis entropy and its application
to black hole physics. In section \ref{FIRST}, we start from the
first law of thermodynamics and derive corrections to the
Friedmann equations through the generalized Kaniadakis entropy. In
section \ref{GSL}, we confirm the validity of the generalized
second law of thermodynamics in this scenario. We finish with
conclusion in the last section.

\section{Kaniadakis horizon entropy \label{Kan}}
In this section, we review the origin and formalism of the
generalized Kaniadakis entropy. Kaniadakis entropy is
one-parameter entropy which generalizes the classical
Boltzmann-Gibbs-Shannon entropy. It originates from a coherent and
self-consistent relativistic statistical theory. The advantages of
Kaniadakis entropy is that it preserves the basic features of
standard statistical theory, and in the limiting case restore it
\cite{Kan1,Kan2}. The general expression of the Kaniadakis entropy
is given by \cite{Kan1,Kan2}
\begin{eqnarray}
S_{K}=- k_{_B} \sum_i n_i\, \ln_{_{\{{\scriptstyle K}\}}}\!n_i  ,
\end{eqnarray}
with $k_{_B}$ is the Boltzmann constant, and
\begin{eqnarray}
\ln_{_{\{{\scriptstyle K}\}}}\!x=\frac{x^{K}-x^{-K}}{2K}.
\end{eqnarray}
Here $K$ is called the Kaniadakis parameter which is a
dimensionless parameter ranges as $-1<K<1$, and measures the
deviation from standard statistical mechanics. In the limiting
case where $K\rightarrow0$, the standard entropy is restored.

In such a generalized statistical theory the distribution function
becomes \cite{Kan1,Kan2}
\begin{eqnarray}
n_i= \alpha \exp_{_{\{{\scriptstyle K}\}}}[-\beta (E_i-\mu)] ,
\end{eqnarray}
 where
\begin{eqnarray}
&& \exp_{_{\{{\scriptstyle K}\}}}(x)=
\left(\sqrt{1+K^2x^2}+K x\right)^{1/K}, \\
&&\alpha=[(1-K)/(1+K)]^{1/2K},\\
&&\beta^{-1}=k_{_{B}}T \sqrt{1-K^2}.
\end{eqnarray}
Let us note that the chemical potential $\mu$ can be fixed by
normalization \cite{Kan1,Kan2}. Alternatively, Kaniadakis entropy
can be expressed as
\cite{Abreu:2016avj,Abreu:2017fhw,Abreu:2017hiy,Abreu:2018mti,Yang:2020ria,
Abreu:2021avp}
\begin{equation}
 \label{kstat}
S_{K} =-k_{_B}\sum^{W}_{i=1}\frac{P^{1+K}_{i}-P^{1-K}_{i}}{2K}.
\end{equation}
Here  $P_i$ is the probability in which the system to be in a
specific microstate and $W$ represents the total number of the
system configurations. Throughout this paper we set
$k_{_B}=c=\hbar=1$.

It is also interesting to apply the  Kaniadakis entropy to the
black hole thermodynamics. It is well known that the entropy of
the black hole, in Einstein gravity, obeys the so called
Bekenstein-Hawking entropy, which states that the entropy of the
black hole horizon is proportional to the area of the horizon,
$S_{BH}= A/(4G)$. Now we assume  $P_i=1/W$, and using the fact
that Boltzmann-Gibbs entropy is $S\propto\ln(W)$, and $S=S_{BH}$,
we get $W=\exp\left[S_{BH}\right]$ \cite{Mor}.

Substituting $P_i=e^{-S_{BH}}$ into Eq. (\ref{kstat}) we arrive at
 \begin{equation} \label{kentropy}
S_{K} = \frac{1}{K}\sinh{(K S_{BH})}.
\end{equation}
When $K\rightarrow 0$ one recovers the standard Bekenstein-Hawking
entropy, $S_{K\rightarrow 0}=S_{BH}$. Considering the fact that
deviation from the standard Bekenstein-Hawking is small, we expect
that $K\ll1$. Therefore, we can expand expression (\ref{kentropy})
as
\begin{equation}\label{kentropy2}
S_{K}=S_{BH}+ \frac{K^2}{6} S_{BH}^3+ {\cal{O}}(K^4).
\end{equation}
The first term is the usual area law of black hole entropy, while
the second term is the leading order Kaniadakis correction term.
In the next section, we shall apply the above expression to
extract the modified friedmann equations.
\section{Corrections to the Friedmann equations\label{FIRST}}
Consider a spatially homogeneous and isotropic spacetime which is
described by the line elements
\begin{equation}
ds^2={h}_{\mu \nu}dx^{\mu}
dx^{\nu}+\tilde{r}^2(d\theta^2+\sin^2\theta d\phi^2),
\end{equation}
where $\tilde{r}=a(t)r$, $x^0=t, x^1=r$, and $h_{\mu \nu}$=diag
$(-1, a^2/(1-kr^2))$ represents the two dimensional subspace. The
parameter $k$ denotes the spatial curvature of the universe with
$k = -1,0, 1$,  corresponds to open, flat, and closed universes,
respectively. The radius of the apparent horizon, which is a
suitable horizon from thermodynamic viewpoint, is given by
\cite{SheyLog}
\begin{equation}
\label{radius}
 \tilde{r}_A=\frac{1}{\sqrt{H^2+k/a^2}}.
\end{equation}
The associated temperature with the apparent horizon is given by
\cite{Cai2,Sheyem}
\begin{equation}\label{T}
T_h=\frac{\kappa}{2\pi}=-\frac{1}{2 \pi \tilde
r_A}\left(1-\frac{\dot {\tilde r}_A}{2H\tilde r_A}\right).
\end{equation}
We also assume the matter/energy content of the universe has the
form of the perfect fluid with energy-momentum tensor,
$T_{\mu\nu}=(\rho+p)u_{\mu}u_{\nu}+pg_{\mu\nu}, $ where $\rho$ and
$p$ are, respectively, the energy density and pressure. The
conservation equation holds for the total matter and energy of the
universe, namely $\nabla_{\mu}T^{\mu\nu}=0$. In the background of
FRW geometry this reads $ \dot{\rho}+3H(\rho+p)=0$, where
$H=\dot{a}/a$ is the Hubble parameter. The work density associated
with the volume change of the universe is defined by \cite{Hay2}
\begin{equation}\label{Work}
W=-\frac{1}{2} T^{\mu\nu}h_{\mu\nu}.
\end{equation}
It is a matter of calculations to show that
\begin{equation}\label{Work2}
W=\frac{1}{2}(\rho-p).
\end{equation}
In order to extract the Friedmann equations from
thermodynamics-gravity conjecture, we assume the first law of
thermodynamics,
\begin{equation}\label{FL}
dE = T_h dS_h + WdV,
\end{equation}
holds on the apparent horizon.  The total energy of the universe
enclosed by the apparent horizon is $E=\rho V$, while $T_{h}$ and
$S_{h}$ are temperature and entropy associated with the apparent
horizon, respectively. One can easily show that
\begin{equation}
\label{dE1}
 dE=4\pi\tilde
 {r}_{A}^{2}\rho d\tilde {r}_{A}+\frac{4\pi}{3}\tilde{r}_{A}^{3}\dot{\rho} dt.
\end{equation}
where $V=\frac{4\pi}{3}\tilde{r}_{A}^{3}$ is the volume enveloped
by a 3-dimensional sphere with the area of apparent horizon
$A=4\pi\tilde{r}_{A}^{2}$. Using the conservation equation, we
find
\begin{equation}
\label{dE2}
 dE=4\pi\tilde
 {r}_{A}^{2}\rho d\tilde {r}_{A}-4\pi H \tilde{r}_{A}^{3}(\rho+p) dt.
\end{equation}
We assume the entropy of the apparent horizon is in the form of
the generalized Kaniadakis entropy (\ref{kentropy2}). In order to
apply entropy (\ref{kentropy2}) to the universe, we need to
replace the horizon radius of the black hole with the radius of
the apparent horizon, namely, $r_{+}\rightarrow \tilde {r}_{A}$.
Therefore, we write the apparent horizon entropy as
\begin{equation}\label{kentropy3}
S_{h}=\mathcal{S}+\frac{K^2}{6} \mathcal{S}^3,
\end{equation}
where $\mathcal{S}=A/(4G)=\pi \tilde{r}_{A}^{2}/G$. Taking
differential form of the Kaniadakis entropy (\ref{kentropy3}), we
get
\begin{eqnarray} \label{dSh}
dS_h&=& d \mathcal{S}+\frac{K^2}{2} \mathcal{S}^2 d \mathcal{S},
\end{eqnarray}
where
\begin{eqnarray} \label{dS}
d \mathcal{S}= \frac{2\pi \tilde{r}_{A}\dot {\tilde r}_A}{G} dt.
\end{eqnarray}
Inserting relations (\ref{Work2}), (\ref{dE2}), (\ref{dSh}) and
(\ref{dS}) in the first law of thermodynamics (\ref{FL}) and using
definition (\ref{T}) for the temperature, after some calculations,
we find the differential form of the Friedmann equation as
\begin{equation} \label{Fried1}
\left(1+\frac{K^2}{2}\mathcal{S}^2\right)\frac{d\tilde
{r}_{A}}{\tilde {r}_{A}^{3}}= 4 \pi G H(\rho+p) dt.
\end{equation}
Using the continuity equation, we reach
\begin{equation} \label{Fried2}
-\frac{2d\tilde {r}_{A}}{\tilde {r}_{A}^{3}} \left(1+\alpha
\tilde{r}_{A}^{4}\right)= \frac{8 \pi G}{3} d \rho,
\end{equation}
where we have defined
\begin{equation}
\alpha\equiv\frac{K^2 \pi^2}{2 G^2}.
\end{equation}
Integrating Eq. (\ref{Fried2}), we arrive at
\begin{equation} \label{Fried3}
\frac{1}{\tilde {r}_{A}^2}-\alpha \tilde {r}_{A}^2  =  \frac{8 \pi
G}{3}\rho+\frac{\Lambda}{3},
\end{equation}
where $\Lambda$ is an integration constant which can be
interpreted as the cosmological constant. Substituting $\tilde
{r}_{A}$ from Eq.(\ref{radius}), we arrive at
\begin{equation} \label{Fried4}
H^2+\frac{k}{a^2}-\alpha\left(H^2+\frac{k}{a^2}\right)^{-1} =
\frac{8 \pi G}{3}(\rho+\rho_{\Lambda}).
\end{equation}
where $\rho_{\Lambda}=\Lambda/(8\pi G)$.  This is the modified
Friedmann equation inspired by the generalized Kaniadakis entropy.
When $\alpha\rightarrow 0$, we find the Friedmann equations in
standard cosmology.

We can also derive the second modified Friedmann equation by
combining the first modified Friedmann equation (\ref{Fried4})
with the continuity equation. If we take the time derivative of
the first Friedmann equation (\ref{Fried4}), after using the
continuity equation, we arrive at
\begin{eqnarray}\label{2Fri2}
\left(\dot{H}- \frac{k}{a^2}\right)\left[1+\alpha
\left(H^2+\frac{k}{a^2}\right)^{-2}\right]=-4\pi G(\rho+p).
\end{eqnarray}
In this way, we derive the modified Friedmann equations given by
Eqs. (\ref{Fried4}) and (\ref{2Fri2}) when the entropy associated
with the apparent horizon is in the form of the generalized
Kaniadakis entropy. Let us note that the modified Friedmann
equations through Kaniadakis entropy was explored in \cite{Lym}.
Our approach in this work has several differences with the one
discussed in \cite{Lym}. First, the authors of \cite{Lym} have
modified the total energy density in the Friedmann equations. The
Friedmann equations derived in \cite{Lym} have the form of the
standard Friedmann equations, with additional dark energy
component that reflects the effects of the corrected entropy.
However, in our approach the modified Kaniadakis entropy affects
the geometry (gravity) part of the cosmological field equations,
and the energy content of the universe does not change. From
physical point of view, our approach is reasonable, since
basically the entropy depends on the geometry of spacetime
(gravity part of the action). As a result, any modification to the
entropy should affect directly the gravity side of the dynamical
field equations. Second, the authors of \cite{Lym} applied the
first law of thermodynamics, $-dE=TdS$, on the apparent horizon
and obtained the modified Friedmann equations through Kaniadakis
entropy. Here $-dE$ is the energy flux crossing the apparent
horizon within an infinitesimal period of time $dt$. While in the
present work, we take the first law of thermodynamics as
$dE=TdS+WdV$, where $dE$ is now the change in the total energy
inside the apparent horizon. Third, the authors of \cite{Lym}
assume the apparent horizon radius $\tilde r_A$ is fixed and
consider the associated temperature as $T=1/(2\pi \tilde r_A)$,
while in this work, due to the cosmic expansion, we assume the
apparent horizon radius changes with time. Therefore, we include
term $WdV$ in the first law of thermodynamics (\ref{FL}).

\section{Generalized Second law of thermodynamics\label{GSL}}
Next we explore the validity of the generalized second law of
thermodynamics when the entropy associated with the horizon is
Kaniadakis entropy (\ref{kentropy3}). In the context of an
accelerating universe, the generalized second law of
thermodynamics were explored in \cite{wang1,wang2,SheyGSL}.

Combining Eq. (\ref{Fried2}) with continuity equation yields
\begin{equation} \label{GSL1}
\dot{\tilde {r}}_{A} (1+\alpha \tilde{r}_{A}^4)= 4\pi G
\tilde{r}_{A}^3 H (\rho+p).
\end{equation}
Solving for $\dot{\tilde {r}}_{A}$, we find
\begin{equation} \label{dotr1}
\dot{\tilde {r}}_{A}=4\pi G \tilde{r}_{A}^3 H (\rho+p) (1+\alpha
\tilde{r}_{A}^4)^{-1}.
\end{equation}
When the dominant energy condition holds, $\rho+p\geq0$, we have
$\dot{\tilde{r}}_{A}\geq0$. We then calculate $T_{h} \dot{S_{h}}$,
\begin{eqnarray}\label{TSh1}
T_{h} \dot{S_{h}}&=&\frac{1}{2\pi \tilde r_A}\left(1-\frac{\dot
{\tilde r}_A}{2H\tilde r_A}\right)\frac{d}{dt}
\left(\mathcal{S}+\frac{K^2}{6} \mathcal{S}^3\right)\nonumber \\
&=& 4\pi G \tilde{r}_{A}^3 H (\rho+p) \left(1-\frac{\dot {\tilde
r}_A}{2H\tilde r_A}\right).
\end{eqnarray}
In an accelerated universe one may have $w=p/\rho<-1$, indicating
the violation of the dominant energy condition, $\rho+p<0$. In
this case, the inequality $\dot{S_{h}}\geq0$ no longer valid and
one should consider the time evolution of the total entropy,
namely the entropy associated with the horizon and the matter
field entropy inside the universe, $S=S_h+S_m$.

The Gibbs equation implies \cite{Pavon2}
\begin{equation}\label{Gib2}
T_m dS_{m}=d(\rho V)+pdV=V d\rho+(\rho+p)dV,
\end{equation}
where the temperature and the entropy of the matter fields inside
the universe are denoted by $T_{m}$ and $S_m$, respectively. We
propose the boundary of the universe is in thermal equilibrium
with the matter field inside the universe. This implies the
temperature of both part are equal $T_m=T_h$ \cite{Pavon2}. If one
relax the local equilibrium hypothesis, then one should observe an
energy flow between the horizon and the bulk fluid, which is not
physically acceptable. From the Gibbs equation (\ref{Gib2}) one
may write
\begin{equation}\label{TSm2}
T_{h} \dot{S_{m}} =4\pi {\tilde{r}_{A}^2}\dot {\tilde
r}_A(\rho+p)-4\pi {\tilde{r}_{A}^3}H(\rho+p).
\end{equation}
Next, we consider the time evolution of the total entropy $S_h +
S_m$. Combining Eqs. (\ref{TSh1}) and (\ref{TSm2}),  we arrive at
\begin{equation}\label{GSL2}
T_{h}( \dot{S_{h}}+\dot{S_{m}})=2\pi{\tilde r_A}^{2}(\rho+p)\dot
{\tilde r}_A.
\end{equation}
Substituting $\dot {\tilde r}_A$ from Eq. (\ref{dotr1}) into
(\ref{GSL2}) we reach
\begin{equation}\label{GSL3}
T_{h}( \dot{S_{h}}+\dot{S_{m}})=8 \pi^2 G H {\tilde
r_A}^{5}(\rho+p)^2 (1+\alpha \tilde{r}_{A}^4)^{-1}.
\end{equation}
In summary, when the horizon entropy has the form of the
generalized Kaniadakis entropy \ref{kentropy3}, the generalized
second law of thermodynamics still holds for a universe enclosed
by the apparent horizon.
\section{conclusion \label{Con}}
It is widely accepted that there is a correspondence between the
laws of gravity and the laws of thermodynamics. This connection
allows to extract the field equations of gravity by starting from
the first law of thermodynamics on the boundary of the system and
vice versa. In this approach the entropy associated with the
boundary of the system plays a crucial role. Any modification to
the entropy expression modifies the field equations of gravity.

In this work, by assuming the entropy associated with the apparent
horizon of the FRW universe is in the form of the generalized
Kaniadakis entropy, and starting from the first law of
thermodynamics, $dE=T_hdS_h+WdV$, we extracted the modified
Friedmann equations describing the evolution of the universe with
any spatial curvature. Since entropy is a geometrical quantity, we
expect any correction to the entropy expression modifies the
gravity (geometry) part of the gravitational field equations.
Therefore, we keep fixed the energy content of the universe, as it
is more reasonable. In the obtained Friedmann equations, the
cosmological constant appears as a constant of integration. We
have also explored the time evolution of the total entropy,
including the entropy of the apparent horizon together with the
entropy of the matter field inside the horizon. We found out that
the total entropy is always a non-decreasing function of time
which confirms that the generalized second law of thermodynamics
holds for the universe with corrected Kaniadakis entropy.

The obtained modified Friedmann equations (\ref{Fried4}) and
(\ref{2Fri2}) provide a background to investigate a new cosmology
based on Kaniadakis entropy. Many issues can be explored in this
direction. One can study the cosmological implications of the
modified Friedmann equations and study the evolution of the
universe from early deceleration to the late time acceleration.
One may also investigate the inflationary scenarios, Big Bang
nucleosynthesis, as well as the growth of perturbations in this
setup. Dark energy scenarios, including the holographic and
agegraphic dark energy models, can be verified based on Kaniadakis
modified Friedmann equations.

\acknowledgments{I highly appreciate the anonymous referees for
constructive comments which helped me improve the paper
significantly. This work is supported by Iran National Science
Foundation (INSF) under grant No. 4022705.}


\end{document}